\documentclass{emulateapj}

\def\bea{\begin{eqnarray}}
\def\ena{\end{eqnarray}}

\usepackage{myaasmacros}

\shorttitle{Deriving structure of the Galactic Magnetic Field }
\shortauthors{Pshirkov et al.}

\begin{document}

\title{Deriving the Global Structure of the Galactic
Magnetic Field from Faraday Rotation Measures of Extragalactic Sources}

\author{M. S. Pshirkov$^{1,2}$, P.G.Tinyakov$^{1,3}$,
P.P.Kronberg$^{4,5,6}$ and  K. J. Newton-McGee$^{6,7}$}

\affil{
$^{1}$Universite Libre de Bruxelles, Service de Physique Theorique, CP225, 1050,  Brussels, Belgium; pshirkov@ulb.ac.be
 \\
$^{2}$Pushchino Radio Astronomy Observatory, 142290,  Pushchino, Russia\\
$^{3}$Institute for Nuclear Research, 117312, Moscow, Russia\\
 $^{4}$Department of Physics, University of Toronto, Toronto M5S 1A7, Canada \\
$^{5}$ IGPP, Los Alamos National Laboratory, M.S. T006, Los Alamos NM 87545, USA\\
$^{6}$ Sydney Institute for Astronomy, School of Physics, The University
of Sydney, NSW 2006, Australia \\
$^{7}$ Australia  Telescope National Facility, CSIRO, PO Box 76, Epping NSW 1710, Australia
}

\begin{abstract}

We made use of the two latest sets of rotational measures (RMs) of
extra-galactic radio sources, namely the NRAO VLA Sky Survey
rotation measures catalog, and a compilation by Kronberg and
Newton-McGee, to infer the global structure of the Galactic
magnetic field (GMF). We have checked that these two data sets are
mutually consistent. Given the existence of clear patterns in
all-sky RM distribution we considered GMF models consisting of two
components: disk (spiral or ring) and halo. The parameters of
these components were determined by fitting different model field
geometries to the observed RMs. We found that the model consisting
of a symmetric (with respect to the Galactic plane) spiral disk
and anti-symmetric halo fits the data best and reproduces the
observed distribution of RMs over the sky very well. We confirm
that ring disk models are disfavored. Our results favor small
pitch angles around $\sim -5^\circ$ and an increased vertical
scale of electron distribution, in agreement with some recent
studies. Based on our fits, we select two benchmark models
suitable for studies of cosmic ray propagation, including the
ultra-high energies.
\end{abstract}



\section{Introduction}
\label{SectionI}
A realistic model of the Galactic magnetic field (GMF) is needed for
various applications, such as direct dark matter searches, studies of
cosmic rays, and others. In particular, the GMF plays a crucial role
in the propagation of ultra-high energy cosmic rays (UHECRs)---those
with energies in excess of $10^{19}$~eV. Their deflections (assuming
these are charged particles) in the GMF are large enough to prevent
the identification of some, perhaps most sources directly from the
UHECR data. However, in the case of protons these deflections are
sufficiently small to be corrected for, provided the GMF is known with
enough accuracy. Thus, the possibility of astronomy with charged
particles at ultra-high energies may depend crucially on our knowledge
of the GMF.

The magnetic field (MF) of our Galaxy is thought to contain both
regular and turbulent components (see, e.g.,
\cite{Beck2001,Beck2008}). While the regular component is
subdominant in strength by a factor of a few, it is expected to
give a dominant contribution in integral quantities such as total
deflections of ultra-high energy particles.  Unlike the turbulent
component, the regular component adds up coherently and comes to
dominate \citep{Tinyakov2005} when integrated over many coherence
lengths. Estimating the strength and global structure of the
regular MF of our Galaxy is the main purpose of this
paper. Our primary interest is on MF outside the thin
Galactic disk, since that is the field that determines the
deflections for most arrival directions of UHECRs.

The regular component of the GMF has been studied previously e.g.
\citep{Simard-Normandin1980,
  Sofue1983,Han1994,Frick2001,Vallee2005,Han2006,
  Brown2007,Sun2008,Jansson2009}, and in particular in the context of
UHECR propagation \citep{Stanev1997, Tinyakov2002, Prouza2003}.
The advance of this paper is that it is based on improved data on
rotation measures (RMs) and the ionized gas distribution. We also
develop improved numerical model commensurate with this improved
data.

Observations of other spiral galaxies reveal MFs
of a few $\mu$G strength, coherent over distances of several kiloparsecs
(see, e.g., \citep{Beck2008,
  Krause2009,Braun2010}).  Observations from the face-on and edge-on
perspectives indicate that these fields consist of at least two
components: the 'disk' one which is concentrated towards the galactic
plane and has largely a spiral pattern, and the 'halo' one that
resides at some distance from the disk plane and has radial size
comparable with the size of the galaxy. As the Milky Way seems to be a
typical member of the entire class of spiral galaxies, MFs
of similar strength and configuration should be expected in our
Galaxy.

There are several methods to estimate the MF in the
Galaxy, such as Zeeman splitting measurements
\citep{Crutcher1999}, infrared polarization studies
\citep{Nishiyama2010}, synchrotron polarization and intensity
surveys
\citep{Wielebinski2005,Sun2008,Jansson2009,Jaffe2010,Ruiz-Granados2010},
starlight polarization studies \citep{Heiles1996} and, finally,
observation of the Faraday RMs of different
Galactic (pulsars) and extragalactic radio sources (e.g.r.s.)
\citep{Simard-Normandin1981,Han2006,Men2008,Mao2010,Kronberg2011}.
Except for the Zeeman measurements all these methods are somewhat
indirect as they involve additional assumptions such as density of
warm ionized medium or concentrations of cosmic ray electrons.  In
this paper we use  recent data on the Faraday RMs.

When propagating through a magnetized plasma, the polarization plane
of a linearly polarized electromagnetic wave of wavelength $\lambda$
rotates by the angle $\Delta \psi$ proportional to the square of the
wavelength,
\begin{equation}
\Delta\psi = \rm{RM}\cdot \lambda^2.
\label{RM1}
\end{equation}
Multi-frequency observations thus allow one to measure the
value of RM. The RM can be expressed in terms of the
properties of the intergalactic/interstellar medium (IGM/ISM) and
permeating MFs:
\begin{equation}
\rm{RM} = 0.81\int_{0}^{D} n_eB_{||} dl,
\label{RM2}
\end{equation}
where $n_e$ is the density of free electrons measured in
cm$^{-3}$, $B_{||}$ is the component of the MF parallel to
the line of sight measured in $\mu$G (positive when directed towards
the observer), and $D$ is the distance from the observer to
the source in pc. Combined with data on the Galactic free electron
distribution, Equations~(\ref{RM1}) and (\ref{RM2}) can be used to estimate
the GMF from the Faraday RMs of a
large number of sources.

Two types of sources are normally used to measure Galactic Faraday
rotation: Galactic pulsars and extragalactic radio sources. Pulsars
have an advantage that their {\em intrinsic} RM values are negligibly
small. However, the use of pulsars requires the knowledge of their
distances (usually inferred from their dispersion measures DM), which
introduces additional assumptions about the ISM density
distribution. In addition, pulsars are concentrated around the
Galactic plane, and their total number ($\sim 500$) is
currently much smaller than the number of extragalactic radio sources
with measured RMs.  For these reasons in this paper, we concentrate on
the RMs of the extragalactic radio sources.

Our general strategy is the following. We consider several types of
analytical GMF models, each containing a finite number of
parameters. For each model we calculate the expected RMs, and fit them to the observations by the binned
$\chi^2$ method. The quality of the fit is then used to determine the
best values of the GMF parameters and their allowed ranges.

The paper is organized as follows. Section \ref{sec:data} describes
the data.  Sections \ref{sec:models} and \ref{sec:method} describe the
implemented models and our method. The results are presented in
Section \ref{sec:results}. Section \ref{sec:conclusions} contains the
results and conclusions.

\section{Data}
\label{sec:data}

\begin{figure}
\begin{center}
\includegraphics[width=80 mm]{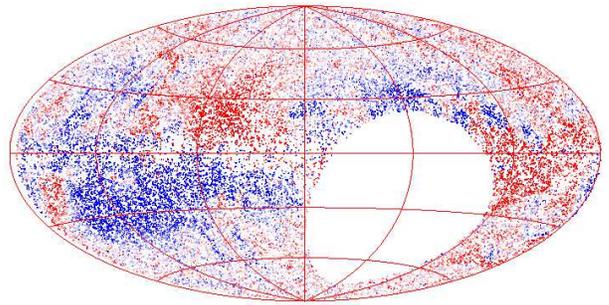}
\end{center}
\caption{Locations and RM signs of the NVSS sources.
Red (blue) color corresponds to
 positive (negative) values of RM.}
\label{figure_nvss}
\end{figure}

In the present work we use two sets of the RM data. The first one is
the compilation of RMs recently obtained by \citet{Taylor2009} who
reanalyzed NRAO VLA Sky Survey (NVSS) data. NVSS is the largest by
number survey of polarized radio sources at declinations
$>-40^{\circ}$ \citep{Condon1998}. It was made in two nearby bands,
1364.9 and 1435.1 MHz; each having a width of 42 MHz.  Simultaneous
observations in these two different bands give estimates of the RMs of
the sources. The total number of the observed sources was 37,543 with
the mean error, estimated by the authors, of $\sim
11~\rm{rad~m^{-2}}$. The sky map of the NVSS\footnote{We henceforth
  refer to NVSS rotation measure catalog as just NVSS.}  set is
shown in Figure~\ref{figure_nvss}, which displays the location and RM
sign of each source.  The data have a large blind spot corresponding
to declinations below $\sim{-40^{\circ}}$.
\begin{figure}
\begin{center}
\includegraphics[width=80 mm]{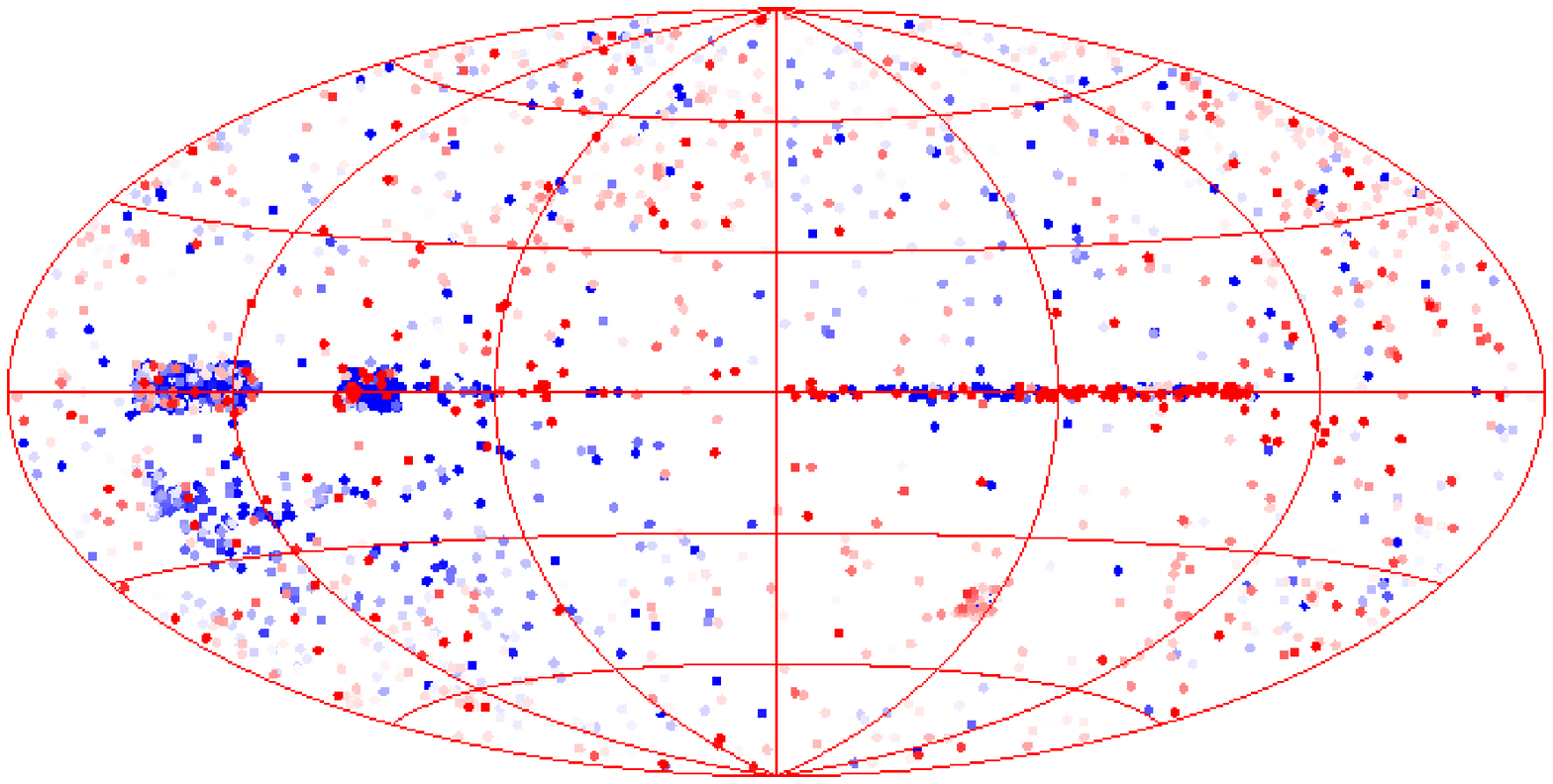}
\end{center}
\begin{center}
\includegraphics[width=80 mm]{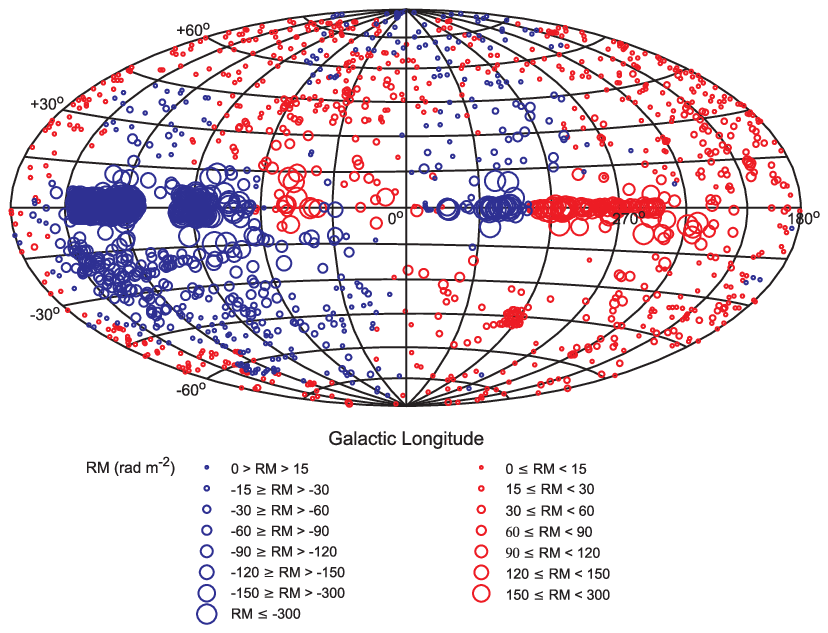}
\end{center}
\caption{Upper panel: locations and RM signs of 2247 sources
  from KNM11.  Red (blue) color corresponds to positive (negative)
  values of RM.   Lower panel: the same data smoothed with a
  resolution of $21^\circ$ (adopted from \citet{Kronberg2011}).}
\label{figure_kronberg}
\end{figure}

An advantage of the NVSS RM catalog is the large number of
sources. However, observations being made in two close bands lead
to relatively large individual RM errors. Also because in general
RM is not always constant in $\lambda^2$, it can have modest
variations over a wider range of $\lambda$. Thus an NVSS RM,
measured in just one small wavelength range at 21 cm ($\lambda=$
20.9 - 22.0 cm), can differ from an RM measured over a wider
wavelength range which tends to average out fluctuations in
RM($\lambda^2$). Such systematic differences will add to the mean
error of $\sim 11~\rm{rad~m^{-2}}$ quoted above by up to $\sim 13$
rad~m$^{-2}$, and will contribute to the widths of the RM
comparison plots with the results from the \cite{Kronberg2011}
compilation (KNM11) shown in Figure~\ref{figure_nvss-vs-Kr} below. On the
other hand, because ${\lambda_2^2-\lambda_1^2}$ is small,
ambiguities of $n~\pi$ in the NVSS RM's are unlikely to occur,
especially at $|b|
> 10^{\circ}$.

It can be easily seen that the NVSS data set should be treated with
care when considering the sources with large observed RMs, which
mainly reside near the Galactic plane and in the region of inner
Galaxy. In order to avoid this problem, we use the NVSS data only at
Galactic latitudes $|b|>10^\circ$.

The second all-sky data set we used is the compilation by
\cite{Kronberg2011} (2257 sources). It consists of $\sim 1500$
revised and statistically more accurate extragalactic radio source
RMs of  P. P. Kronberg et al. (in preparation). These are derived from
polarization measurements made over a large ``baseline''of
$\lambda^2$. The remaining RM's are taken from the Canadian
Galactic Plane Survey  \citep{Brown2003}, the Southern
Galactic Plane Survey  \citep{Brown2007}, and two smaller
published lists from \citet{Klein2003} (108 RMs) and
\citep{Mao2008} (68 RMs). The extragalactic radio sources in the
KNM11 set were carefully observed at multiple frequencies and give
RMs largely free from ambiguities. Errors of the RM
measurements in the KNM11 set are typically of order
$\sim3~\rm{rad\,m^{-2}}$.  We use this set for control.

The sky map of the KNM11 RMs is shown in
Figure~\ref{figure_kronberg}, upper panel. The lower panel shows the
same data smoothed over the circle of $21^\circ$ centered at the
location of each source as described by \citet{Kronberg2011}.  One
may observe a generally good agreement in the RM sky distributions
of the NVSS and KNM11. This is particularly visible when comparing
Figure~\ref{figure_nvss} with the lower panel of
Figure~\ref{figure_kronberg}.

The RM sky structure is rather peculiar. The regions of positive and
negative RMs are roughly symmetric with respect to the
Galactic plane in the outer Galaxy $90^\circ < l < 270^\circ$ and
anti-symmetric in the inner Galaxy. Note, however, that the boundary
between positive and negative RMs in the inner Galaxy is displaced
away from the Galactic plane (toward negative $b$ for
$0<l<90^\circ$),that is, RMs do not change sign across the Galactic
plane itself. This fact was observed, e.g., in
\cite{Kronberg2011}. This overall structure of RMs suggests,
qualitatively, a Galactic field structure that consists of two
components, with the disk field being symmetric in $b$ with respect to
the Galactic plane and the halo field anti-symmetric.

Although the second data set is significantly smaller, it has much
smaller individual RM errors and can be used in two ways.  First, it
allows for an independent evaluation of the quality of the NVSS
data. In order to assess the accuracy of the NVSS RMs we
identified, by positional cross-correlation, the sources that are
common to both catalogs. We did it separately for the two regions
$|b|>10^\circ$ and $|b|<10^\circ$. We found 1338 pairs and 306 pairs
of sources, respectively, separated by less than $0^\circ.05$. For
these sources, we built distributions of differences between RMs given in the NVSS catalog and in the set by KNM11. The
difference distributions are shown in Figure~\ref{figure_nvss-vs-Kr}.
\begin{figure}
\begin{center}
\includegraphics[width=8cm]{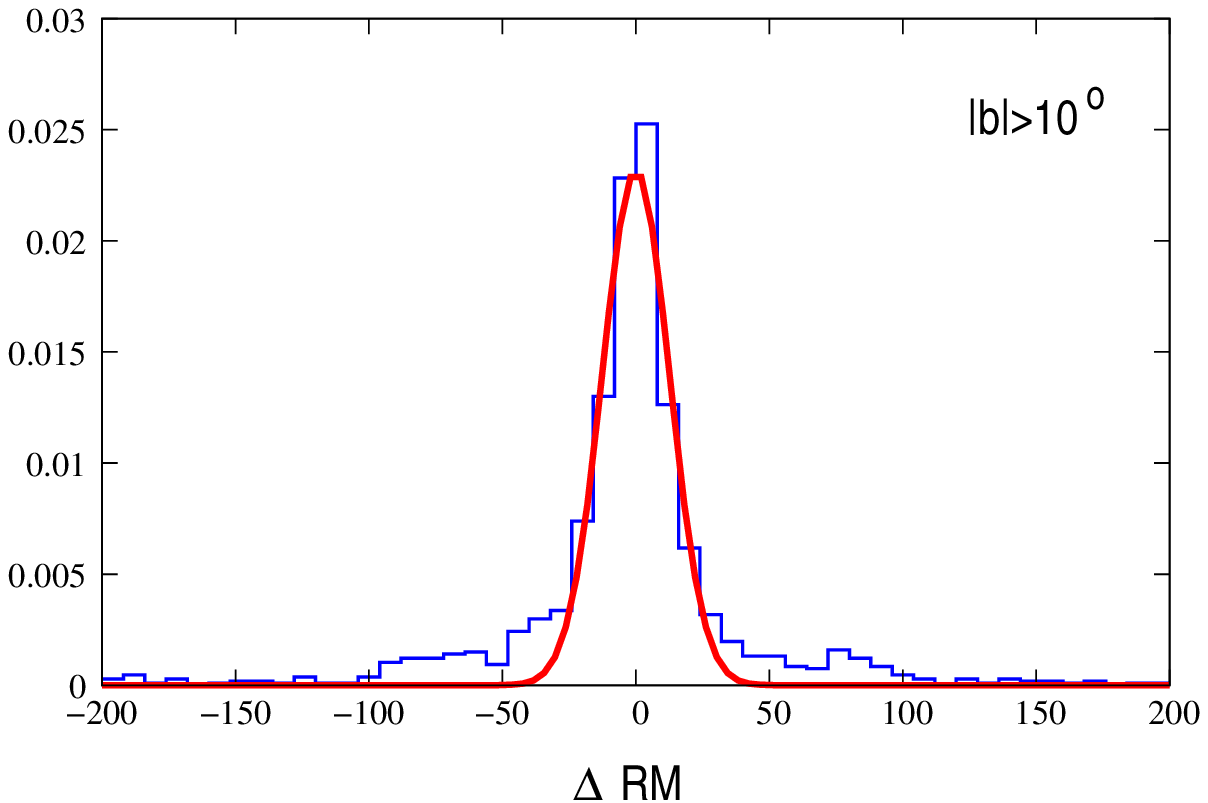}
\includegraphics[width=8cm]{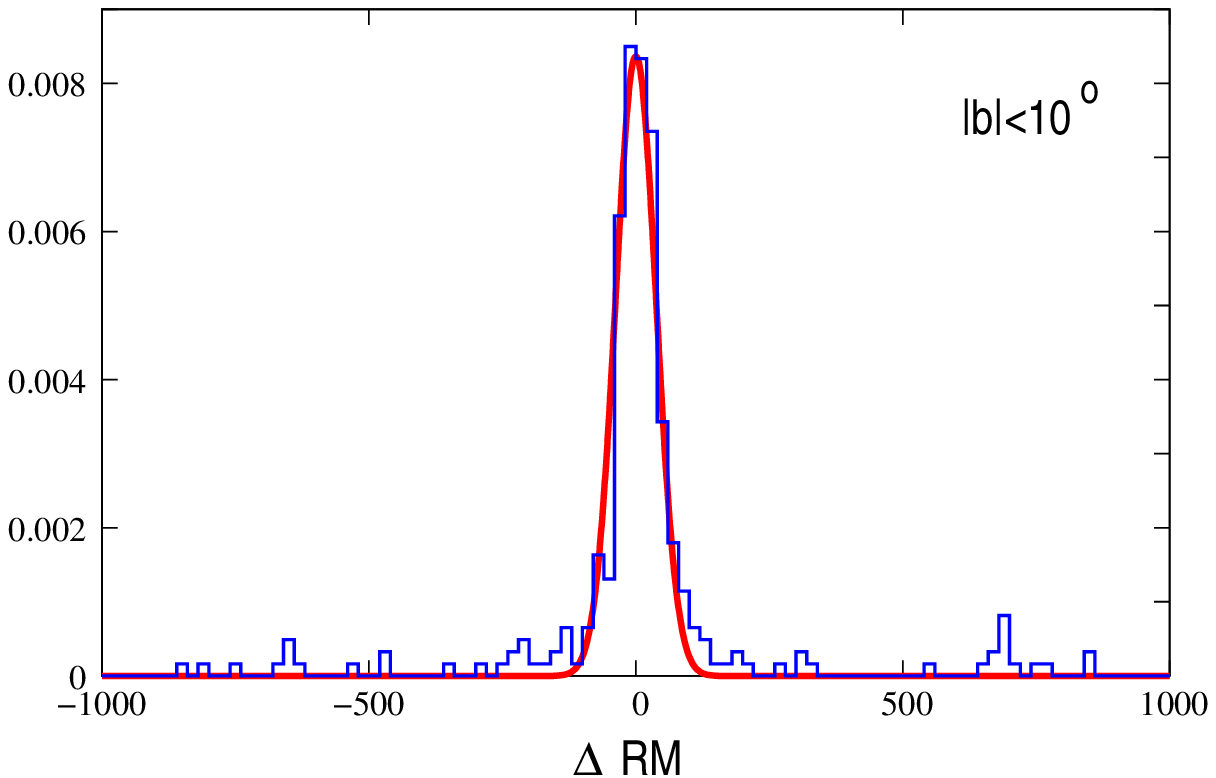}
\end{center}
\caption{Distribution of differences of RMs as given by the NVSS
 catalog and by KNM11. The upper curve, for $|b| > 10^{\circ}$,
shows a Gaussian fit for the range (--200, 200) rad m$^{-2}$. The
lower curve, for $|b| < 10^{\circ}$, shows the corresponding
result for the full range (--1000, 1000) rad m$^{-2}$.
}
\label{figure_nvss-vs-Kr}
\end{figure}

In both regions, the distributions of $\Delta {\rm RM}$ have peaks
centered at zero and well fitted by Gaussians with  dispersions
$\sigma\simeq 13~ \rm{rad~m^{-2}}$ and $\sigma\simeq
36~\rm{rad~m^{-2}}$, respectively. In the region $|b|>10^\circ$
the tails contain $\sim 25$\% of points and are confined within
$\pm 100~\rm{rad\,m^{-2}}$. The effective error introduced by
these tails is of the order of the Gaussian part. Thus, in this
region the NVSS data can be used for our purposes.

On the contrary, in the region $|b|<10^\circ$ the tails
(containing similar fraction of points) are spread over a very
wide range $\pm 900~\rm{rad\,m^{-2}}$. Such tails would introduce
errors in the low-$b$ bins which are too large to constrain the
fit. For this reason we exclude the  bins with $|b|<10^\circ$ from
the NVSS fit, as will be described in Section~\ref{sec:method}.

The KNM11 RM set provides statistically more accurate RMs
 in the Galactic plane, including areas where the NVSS
data do not give sky coverage.  This allows us to check whether
the models which fit the NVSS data also fit the RM data in the
Galactic plane.

\section{GMF and electron density models}
\label{sec:models}

\begin{figure}
\begin{center}
\includegraphics[width=8cm]{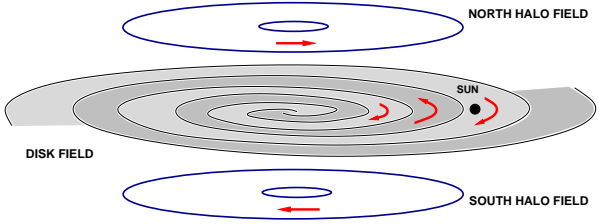}
\end{center}
\caption{Sketch of the  structure of the galactic magnetic field.
The disk field in this sketch has reversals of the MF in adjacent
arms. It is shown that the halo field is antisymmetric with
respect to the galactic plane: in the solar neighborhood the
direction of the halo field is anti-coincident with the direction
of the disk field above the galactic plane and coincident below.
We assumed that the disk field is symmetric with respect to the
galactic plane, in the sense that the sign of the disk field is
locally preserved across $b=0$.}

\label{3Dmodel} \label{sketch_model}
\end{figure}

We adopt a general model of the GMF consisting of two different
components: a disk field and a halo field
\citep{Prouza2003,Sun2008}. Each of these two components is
parameterized independently. According to the qualitative features of
the RM distribution discussed above, for the most part of this work we
take the disk field to be symmetric with respect to the Galactic plane
and the halo field anti-symmetric (other possibilities will be
discussed in Section~\ref{sec:results}).  The general layout of the
model is shown in Figure~\ref{sketch_model}.
\\~ \\

\subsection{The disk field}
\label{sec:disk-field}

We consider several disk GMF models. The first one is the widely
used logarithmic spiral model, e.g.,
\citep{Simard-Normandin1980,Han1994,Stanev1997,Tinyakov2002}.
There are two versions of this model depending on whether the
direction of the field in two different arms is the
same\footnote{In the work of
  \cite{Jansson2009} this model is dubbed as DSS-\textit{Disymmetric
    Spiral}.}  (axisymmetric, or ASS model) or opposite (bisymmetric, or BSS model). They are defined
as follows:
\[
B_{\theta} = B\cos{p}, ~B_{r} = B\sin{p},
\]
\begin{equation}
B(r,\theta, z)=B(r)\left|\cos{\left(\theta-
b\ln{\frac{r}{R_{\odot}}} +\phi\right)}\right|\exp{(-|z|/z_0)},
 \label{MFASS}
\end{equation}
or
\begin{equation}
 B(r,\theta,z)=B(r)\cos{\left(\theta- b \ln{\frac{r}{R_{\odot}}}
+\phi\right)}\exp{(-|z|/z_0)},
\label{MFBSS}
\end{equation}

for ASS and BSS models, respectively. Here
\[
\phi=b \ln\left(1+\frac{d}{R_{\odot}}\right)-\frac{\pi}{2},
\]
\[b\equiv 1/\tan p,
\]
where $p$ is the pitch angle and $d$ is the distance to the first
field reversal. Negative $d$ means that the nearest reversal occurs in
the direction to the Galactic center, positive corresponds to the
opposite direction. $R_{\odot}$ is the distance to the Galactic Center
(we adopt $R_\odot=8.5$~kpc in this paper). The amplitude of the GMF
is a function of the radial coordinate $r$,
\begin{equation}
B(r)= \left\{
\begin{array}{ll}
\displaystyle B_0\frac{R_{\odot}}{r\cos{\phi}}, \quad &r>R_c,\\
\displaystyle B_0\frac{R_{\odot}}{R_c\cos{\phi}}, & r<R_c,
\end{array}
\right.
\end{equation}
where $R_c$ is the radius of the central region where the disk field is
assumed to have constant magnitude. In the course of the simulations, the
parameters were varied within the following ranges:
\begin{center}
\begin{tabular}{l|l|l}
parameter & min & max \\ \tableline
$R_c$ & $3$~kpc & $6$~kpc\\
$z_0$ & $0.5$~kpc & $1.5$~kpc\\
$d$ & $-1.4$~kpc & $1.4$~kpc\\
$p$ & $-15^\circ$ & $15^\circ$ \\

\tableline
\end{tabular}
\end{center}

The second model of the disk field was the axisymmetric field with
reversals in concentric rings. It
was taken from \citet{Sun2008}:
\[
B(r,\theta,z)=D_1(r,z)D_2(r)
\]
\begin{equation}
D_1(r,z)=\left \{
\begin{array}{ll}
\displaystyle B_0\exp\left(-\frac{r-R_{\odot}}{R_0}
-\frac{|z|}{z_0}\right),\; & r>R_c\\
\displaystyle B_0\exp\left(-\frac{|z|}{z_0}\right), &r\leq R_c
\end{array}
\right.
\end{equation}
\begin{equation}
D_2(r)= \left\{
\begin{array}{ll}
+1,\quad &r>7.5~\mathrm{kpc},\\
-1,&6~\mathrm{kpc}<r\leq7.5~\mathrm{kpc},\\
+1,&5~\mathrm{kpc}<r\leq6~\mathrm{kpc},\\
-1,&r\leq5~\mathrm{kpc},
\end{array}
\right.
\label{eq:D2}
\end{equation}
where $R_0$ is  radial scale of the MF.
We studied the following parameter space:
\begin{center}
\begin{tabular}{l|l|l}
parameter & min & max \\ \tableline
$R_c$ & $4.5$~kpc & $5.5$~kpc\\
$R_0$ & $8$~kpc & $15$~kpc\\
$z_0$ & $0.5$~kpc & $1.5$~kpc\\
\tableline
\end{tabular}
\end{center}
We also varied slightly the radial positions of the inner MF
reversals (the boundaries of regions of different sign in
Equation~(\ref{eq:D2})) in the ranges $4.8--5.2$~kpc, $5.8--6.2$~kpc and
$7.3--8.0$~kpc.

The magnitude of the disk MF at the vicinity of the
solar System $B_0$ was taken to be $2~\mu$G. Note that, in the
total GMF, the disk and halo components will be combined with
arbitrary coefficients (to be determined by fitting), so the
choice $B_0=2~\mu$G at this point is merely a matter of
definition.

\subsection{The halo model}

The basic halo model was taken from \citep{Prouza2003,Sun2008}:
\[
B_{\theta}^H(r,z)=
\]
\begin{equation}
B_0^H\left[1+\left(
\frac{|z|-z_0^H}{z_1^H}\right)^2\right]^{-1}\frac{r}{R_0^H}
\exp{\left( 1-  {r\over R_0^H} \right)},
\label{halo1}
\end{equation}
where the direction of the field is reversed in the Southern
hemisphere \citep{Brandenburg1992,Han1997}. We repeat here the
following definitions from  \cite{Sun2008}: $B^H_0$ is the halo MF
strength, $R^H_0$ is its radial scale,  parameter $z^H_0$ defines
the vertical position of the halo,  $z_1^H$ is its vertical scale;
this scale could differ  in  directions to and away from the
galactic plane, $|z|<z_0^H$ and $|z|>z_0^H$ respectively. We will
refer to these parameters as $z^H_{1(1)}$ and $z^H_{1(2)}$,
correspondingly. Here and below the index "H" denotes parameters
belonging to the halo. The original values of the parameters in
the model of \cite{Sun2008} were: $B^H_0 = 10 ~\mu$G, $R^H_0 = 4$
kpc, $z_0^H = 1.5~\mathrm{kpc}$ and $z_1^H = 0.2~\mathrm{kpc}$ for
$|z|<z_0^H$, otherwise $z_1^H = 0.4~\mathrm{kpc}$ ( $z^H_{1(1)} =
0.2~\mathrm{kpc}$ and $z^H_{1(2)} = 0.4~\mathrm{kpc}$).  We also
tested halo models with slightly different radial profiles:
\[
B_{\theta}^H(r,z)=
\]
\begin{equation}
B_0^H \left[ 1+\left(\frac{|z|
-z_0^H}{z_1^H}\right)^2\right]^{-1}
\exp{\left(-\left(\frac{r
-R_0^H}{R_0^H}\right)^2\right)},
\label{halo2}
\end{equation}
and
\[
B_{\theta}^H(r,z)=
\]
\begin{equation}
B_0^H  \left[ 1+\left(\frac{|z|-
z_0^H}{z_1^H}\right)^2\right]^{-1}
\exp{\left(-\left|\frac{r
-R_0^H}{R_T}\right|\right)}.
\label{halo3}
\end{equation}
We investigated the following parameter space:
\begin{center}
\begin{tabular}{l|l|l}
parameter & min & max \\ \tableline
$R_0^H$ & $3.5$~kpc & $15$~kpc\\
$R_T$ & $1$~kpc & $5$~kpc\\
$z_0^H$ & $0.8$~kpc & $3.5$~kpc\\
$z_{1(1)}^H$ & $0.2$~kpc & $0.4$~kpc\\
$z_{1(2)}^H$ & $0.3$~kpc & $0.5$~kpc\\
\tableline
\end{tabular}
\end{center}
Given that the halo field has not been firmly established
as entirely azimuthal, we further studied the basic halo model,
varying the pitch angle within the range $-10^{\circ}<p^H<10^{\circ}$.

\subsection{The electron density}
\label{sec:electron-density}

Calculation of the RMs, Equation~(\ref{RM2}), requires a
knowledge of the Galactic electron density $n_e$. In this paper we
adopted the NE2001 model developed by \citet{Cordes2002}. However,
there is mounting evidence that the original vertical scale of
this model is insufficient and pulsar observations are better
described by the electron density with larger vertical scale
\citep{Gaensler2008}. We tried both the original
($z_h=0.95~\mathrm{kpc}$, $n_e=0.03~\mathrm{cm^{-3}}$ at $z=0$)
and modified versions ($z_h=1.8~\mathrm{kpc}$,
$n_e=0.014~\mathrm{cm^{-3}}$ at $z=0$) of NE2001.

The preliminary simulations with both versions of NE2001 were made
with a coarse grid of GMF parameters. The original model performed
systematically worse than the modified one, so we proceeded to the
full simulation with the modified version of NE2001.  The results
presented below were obtained with this modified version. Apart
from the test of two vertical scales as described in
Section~\ref{sec:results}, we did not vary any other parameters of
NE2001 during the fits.

\section{Method}
\label{sec:method}

\subsection{Binning}

\begin{figure}
\begin{center}
\includegraphics[width=8cm]{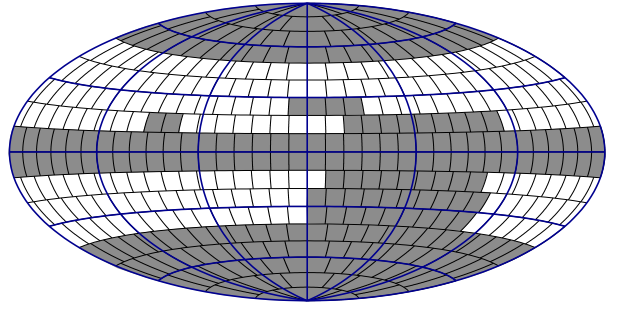}
\end{center}
\caption{Binned celestial sphere. The bins marked in gray are excluded
from the fit (see Appendix~\ref{ExcludedBins} for details). }
\label{figure_bin}
\end{figure}

\begin{figure}
\begin{center}
\includegraphics[width=7.5cm]{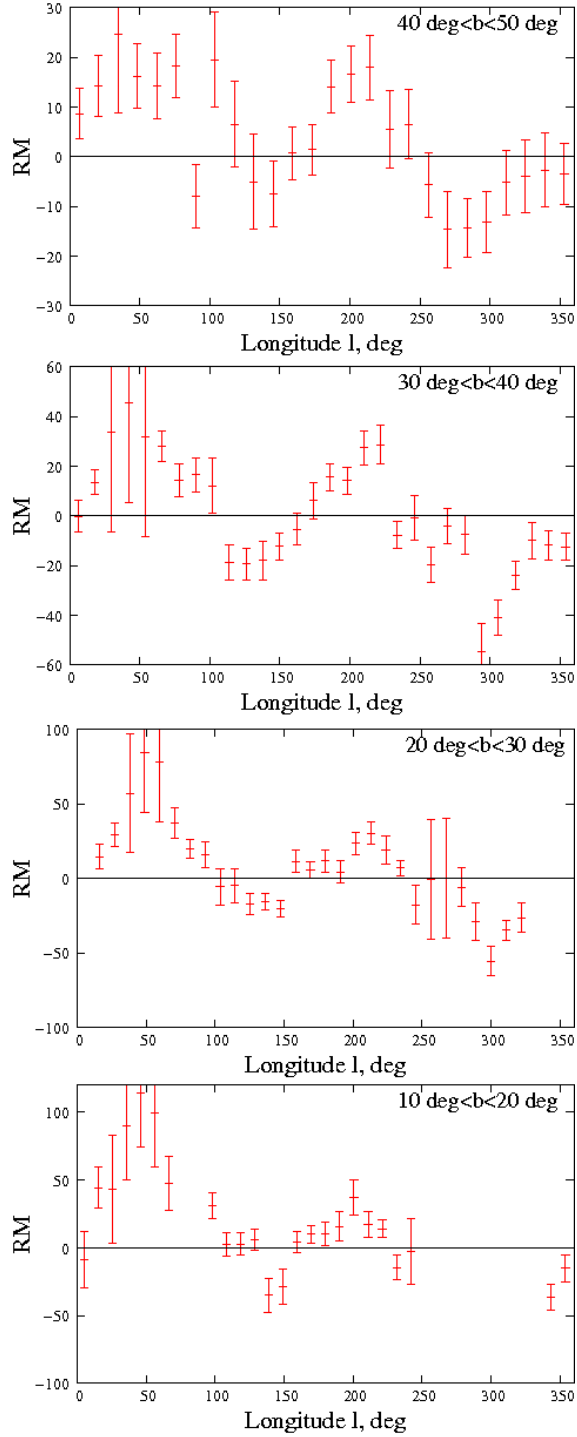}
\end{center}
\caption{Averaged RMs in four bands $10^{\circ}<b<50^{\circ}$. The
  error bars show the errors used in the fit. }
\label{figure_bands_north}
\end{figure}

\begin{figure}
\begin{center}
\includegraphics[width=7.5cm]{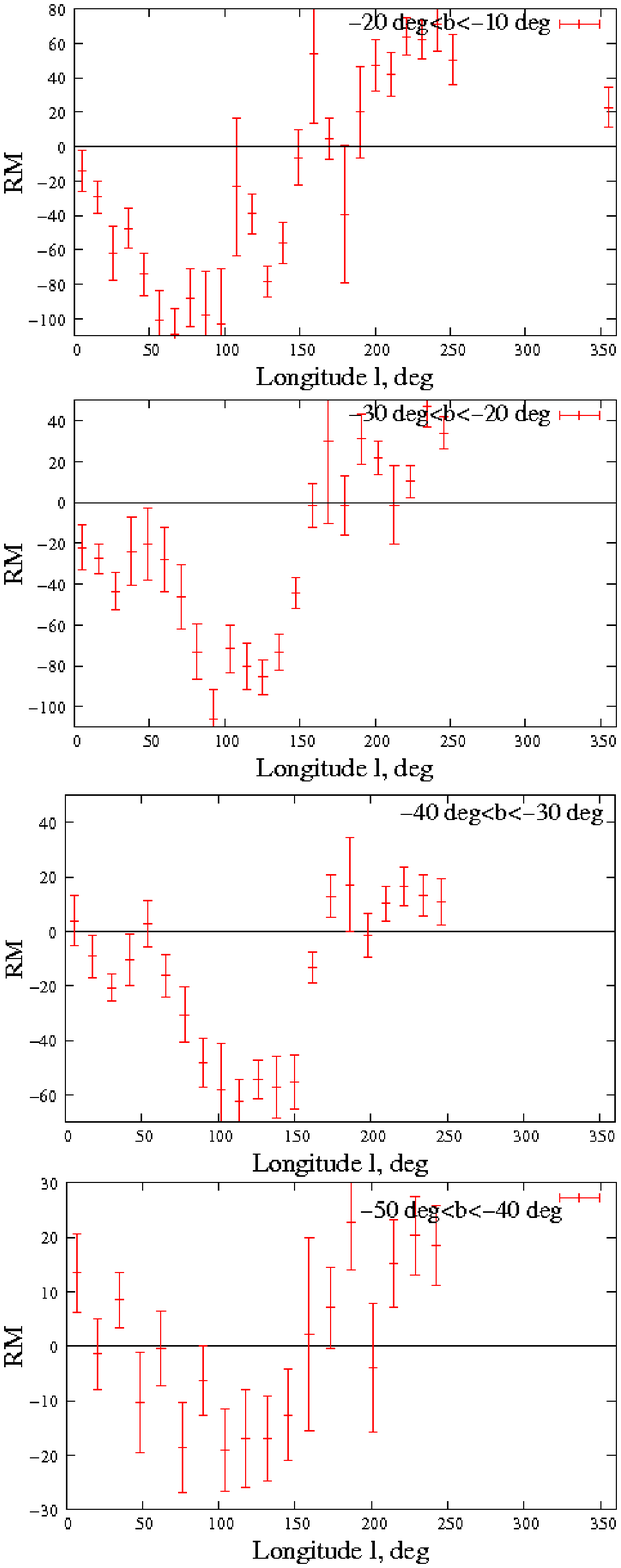}
\end{center}
\caption{Averaged RMs in four bands $-50^{\circ}<b<-10^{\circ}$. The
  error bars show the errors used in the fit.}
\label{figure_bands_south}
\end{figure}

RMs of individual sources are affected by
uncertainties due to the intrinsic source contributions to the RM
and by small-scale fluctuations of the GMF. The effect of the
coherent component of the GMF can be revealed by averaging over a
large number of nearby sources. The angular bin size should be
small enough so that variations of the coherent component of GMF
within the bin can be neglected, and it should include a
sufficient number of sources to suppress source-to-source
fluctuations. Applying these requirements, we chose the following
binning: the celestial sphere is divided into 18 bands of
10$^{\circ}$ width, from --90$^{\circ}$ to +90$^{\circ}$ in
Galactic latitude $b$.  Each band is further divided into bins so
that individual bins would have roughly equal area, about 100
deg$^2$, as shown in Figure~\ref{figure_bin}.
So, there are 36 bins in the 9th band ($0^{\circ}<b<10^{\circ}$), but
only four in the first one ($80^{\circ}<b<90^{\circ}$). The total number
of bins is equal to 422. After averaging, all the NVSS data are
reduced to 422 values of average RM in each bin. No outliers were
removed at this stage.  Figures~\ref{figure_bands_north}, and 
\ref{figure_bands_south} shows the averaged RMs in the eight bands
with $10^{\circ}<b<50^{\circ}$ and $-50^{\circ}<b<-10^{\circ}$,
respectively. The coherent structure is clearly visible.

The GMF models were treated in a similar way. For each model we
simulated 20,000 values\footnote{We checked that increase in
  number of simulated sources from 20,000 to 40,000 does not lead to
  any significant changes.} of RMs at randomly chosen locations and then
applied the same binning and averaging procedure to these sets. The
resulting averaged RMs were compared to observations.

A somewhat different procedure was used for the set of sources
from KNM11. To check the disk field, we used the sources residing
within $1^{\circ}$ from the galactic plane. As their number is more
limited (about 240) and they are located irregularly within their
respective bins (10 deg$^2$ in this case), we simulated model RM
values in their exact locations to avoid an additional error that
would be produced by different distributions of simulated and real
sources within the bin.

\subsection{Errors}
\label{sec:errors}

Important for this analysis is the weighting of different bins or,
equivalently, the bin-by-bin estimation of errors. There are
several sources of scatter of individual RMs within the bin:
internal RM of the source, an anomalous extragalactic RM
component, random component of the Galactic field, variation of
the regular GMF over the bin, fluctuations of the Galactic
electron density $n_e$, and the measurement error. All these
factors contribute to deviations of the observed average value of
RM from that predicted by the GMF model, and determine the error
that has to be assigned to the bin.

If all the above factors were random, it would be possible to
estimate the error by statistical means. Note, however, that some
of these factors (e.g., random component of GMF with coherence
length comparable to the bin size, fluctuations of $n_e$) may act
coherently over the bin. These factors are much more difficult to
treat rigorously. We therefore take a phenomenological approach
and assume that these factors produce an error in the bin that is
some fraction of the root mean square (rms) of the RMs in that bin. In Appendix~\ref{ExcludedBins} we develop a
toy model where such proportionality takes place.

Taking account of this complication, we have estimated the errors of
RMs in the individual bins as follows. By considering the sources in
the directions of north and south Galactic poles where the effect of
the regular MF is expected to be suppressed, the
dispersion in the observed RMs of quasars is estimated as $\sigma_{\rm
  qso}=15$~rad$\cdot$m$^{-2}$ (see Appendix~\ref{ExcludedBins}). If
the rms $\sigma_i$ of RMs in the $i$th bin did not exceed this
value, we set the error of average RMs to $\delta_i=\sigma_{\rm
  qso}/\sqrt{N_i}$, where $N_i$ is the number of sources in the
bin. The reason is that in this case there is no indication of
coherence effects, so one may assume that the error is due to
stochastic factors such as the internal scatter, measurement error,
and short-scale random MFs.

In the opposite case we assume that the variance in the observed RMs
is (partly) due to the presence of some unknown features that could
act coherently over the bin, but cannot be accurately modeled. We
assigned the error of $\sigma_{i}/3$ to these bins, where the factor
$1/3$ is taken from the toy model of Appendix~\ref{ExcludedBins}.  We
have checked that the best-fit parameters do not change significantly
when this factor is varied in the range $1/2-1/4$. This uncertainty in
the absolute normalization of errors we cannot reduce further.

After a bin-by-bin examination, we have increased the errors in
bins where there are clearly visible anomalies in the observed
data (see the list of these bins below in the Appendix
\ref{ExcludedBins}). These problems are: (1) existence of small
cluster of sources with RMs sharply different from the surrounding
values (2) a strongly irregular distribution of the observed
sources across the bin (due to, e.g., deep surveys within small
regions) that could bias the average compared to the case of a
random distribution of the simulated sources, and (3) the presence of
sources with absolute values of RMs larger than 300 rad m$^{-2}$
which should be treated with suspicion due to the $n\pi$ ambiguity
in the definition of polarization angle. The total number of bins
with errors increased by hand is about 10\%. The algorithm is
described in Appendix \ref{ExcludedBins}.

Some bins were completely removed from the analysis.  First, these are
the bins in polar regions with $b>50^{\circ}$ (50 bins) and
$b<-40^{\circ}$ (77 bins). In these regions the model values of RM are
small because the projection of the MF onto the line of
sight is suppressed (recall that in our model both disk and halo
fields are parallel to the Galactic plane regardless of the
parameters).  On the other hand, there is strong evidence that some
sort of 'screen' is present in the direction of high latitudes
\citep{Stil2011} which may provide a dominant contribution in these
regions. The averaged RMs in polar regions, together with the assigned
errors, are shown in Figures ~\ref{northern_cap}, and \ref{southern_cap}
where some bias toward positive values of RMs can indeed be seen.

Second, all bins with $|b|\leq10^{\circ}$ were excluded because we
could not trust the accuracy of the NVSS values in this region as
explained in Section~\ref{sec:data}. Finally, we excluded all bins
where the number of observed sources was smaller than 30 (this
primarily concerns bins overlapping with the blind spot).
Figure~\ref{figure_bin} shows the excluded bins in gray.

A similar prescription was adopted for the analysis of the 1$^{\circ}$
strip along the Galactic plane using the RM set by KNM11, with the
following changes. As the area in this case is 10 times smaller
($\sim 10$~deg$^2$ instead of $100$~deg$^2$) we set our lower limit to
three sources in the bin. Also, we increased the assigned error value to
$\sigma_{i}$ instead of $\sigma_{i}/3$, thus attempting to
reproduce the enhanced uncertainty in the disk due to the presence of
a large number of local structures and very long propagation path
length inside the disk (of order of tens of kpc). In the 1$^{\circ}$
strip, after visual examination, we removed 20 sources out of total
number of 259. RM values of the removed sources clearly demonstrate
large differences with RMs of neighboring sources.

\subsection{Fitting procedure}
\label{sec:fitting-procedure}

The best values of the GMF parameters were identified by the maximum
likelihood method. This amounts to minimizing the value of the
$\chi^2$ defined as follows,
\begin{equation}
\chi^2 = \sum_{i=1}^{\rm Ntot} \frac{(\rm{RM}_{\rm obs}-\rm{RM}_{\rm
    sim})_{i}^2}{\delta_{i}^2},
\label{chi2}
\end{equation}
where $N_{\rm tot}$ is the total number of bins that we have included
in the fit, $(\rm{RM}_{\rm obs}-\rm{RM}_{\rm sim})_i$ is the difference between
the observed and simulated average RM values for the $i-$th bin and
$\delta_{i}$ is the adopted error of the observed average.

Since the RMs (Equation~(\ref{RM2})) are linear in the
MF, the total RM of a given source is a simple sum of
the disk and halo contributions.  Making use of this property, we
separately simulated more than 30,000 halo RM maps and 10,000 disk
RM maps, corresponding to different parameters varied within the
ranges described in Section~\ref{sec:models}. At this stage the
strengths of the disk and halo components were held fixed and set
to values given in Section~\ref{sec:models}. Then each disk
configuration was combined with each halo configuration with two
arbitrary coefficients $\alpha$ and $\beta$.  For a global GMF
model thus obtained, the value of $\chi^2$ in Equation~(\ref{chi2}) is
a quadratic polynomial in $\alpha$ and $\beta$, whose minimum with
respect to $\alpha$ and $\beta$ we found analytically. The
resulting values of $\alpha$ and $\beta$ determined the strength
of the disk and halo fields.

The parameters of the GMF field were varied, within the ranges given
in Section~\ref{sec:models}, by two different methods: by uniformly
covering a given parameter range with a regular grid (grid-type
simulation) and by randomly picking parameter values from a uniform
distribution over the desired range (Monte Carlo type
simulations). The disk field was simulated using the grid method,
whereas for the halo we used both methods in roughly equal proportion.

Because of the uncertainty in the absolute normalization of
$\chi^2$ resulting from uncertainties in the error estimation (see
Section~\ref{sec:errors}), we did not combine the Northern and
Southern Galactic hemispheres in a single fit. The two hemispheres
do not come on equal footing because of the blind spot in the
Southern sky. Instead, our general fitting strategy was as
follows. First we performed the fit using the NVSS data for the
Northern hemisphere only, thus obtaining the preliminary set of
models. Next, we fitted the NVSS data in the Southern hemisphere.
We chose only the models that performed well in both tests and had
the same value of the disk field. Note that this allowed for
different halo fields in the Northern and Southern hemispheres. In
these tests we imposed the additional constraint that
$|\textbf{B}|$ be greater than 1.8 $\mu$G in both hemispheres, to
conform with the results of other observations. Models were
selected as acceptable if they satisfied the following
requirement,
\begin{equation}
\chi^2/\chi^2_{\rm min}=\chi^2_{\rm red}/\chi^2_{\rm red, min}<1+\sqrt{{\rm 2/N_{\rm d.o.f.}}},
\label{eq:chi2}
\end{equation}
where $\chi^2_{\rm min}$ ($\chi^2_{\rm red, min}$) is a minimal
value of $\chi^2$ ($\chi^2_{\rm red}$). These are the models
having parameters within $\sim 1 \sigma$ from the best-fit ones.
Note that the criterion (\ref{eq:chi2}) is insensitive to the
absolute normalization of the errors. We choose such a criterion
to reduce the effect of the ambiguities in the error estimation
discussed in Section~\ref{sec:errors}.

Independently, we fitted the sources from the compilation by KNM11
in the narrow strip of the galactic plane $|b|<1^{\circ}$. This
imposed additional constraints on the structure of the field in
the thin disk to which our fit of NVSS data is insensitive.
Requiring that fits to both the NVSS and KNM11 data be good, we
performed the final selection of the models.

Overall, we required that a successful model passes three separate
fits: to the NVSS data in the Northern and Southern hemispheres,
and to the KNM11 data in the disk. Note that this is, in general, more
restrictive than performing a single combined fit.

\section{Results}
\label{sec:results}

\begin{figure}
\begin{center}
\includegraphics[width=8cm]{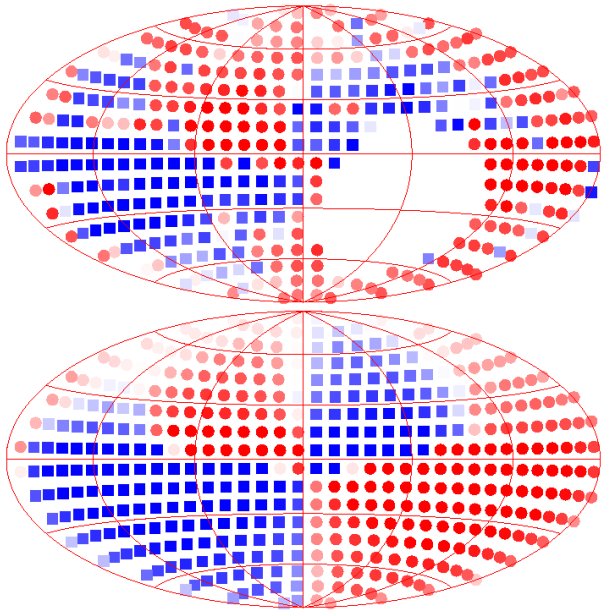}
\end{center}
\caption{Average rotation measures in bins. Red circles (blue squares)
  represent positive (negative) RMs. The color intensity reflects the
  absolute magnitude.  Top: NVSS data. Bottom: best-fit model. }
\label{fig:bin_nvss}
\end{figure}

Figure~\ref{fig:bin_nvss} shows a comparison between one of the best fit
models and the binned NVSS data. Positive average RMs are shown by red
circles, and negative by blue squares. The intensity of the color
indicates the absolute value of RM.  It is seen that the general
structure of the field is reproduced quite well.

\begin{figure}
\begin{center}
\includegraphics[width=8cm]{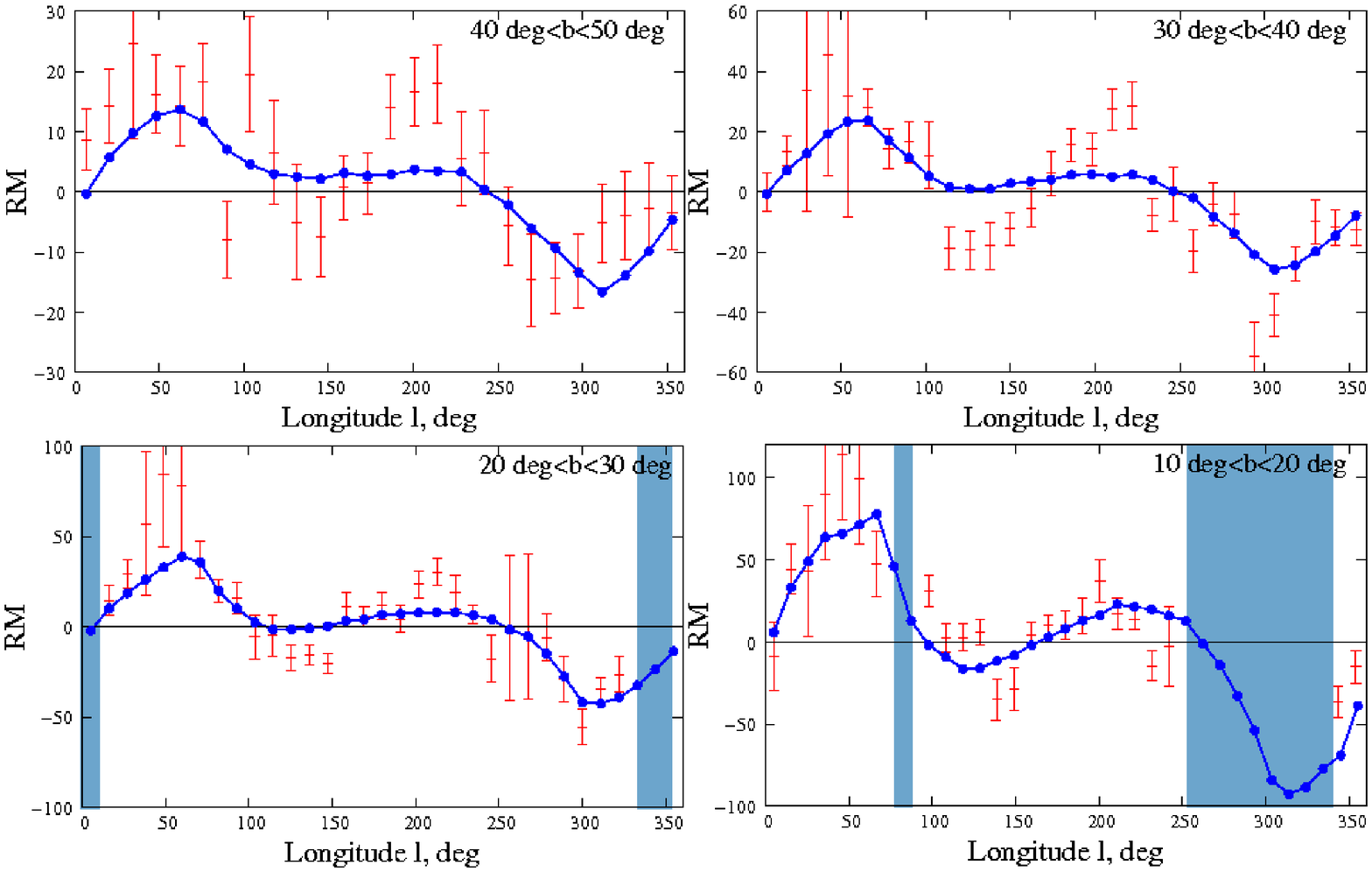}
\end{center}
\caption{Best-fit model (blue solid line) vs. NVSS data (red points)
in the Northern hemisphere. Bins not included in the fit are shaded
with light blue.}
\label{north_fit}
\end{figure}

The results of the fits are as follows:

\noindent {\it North}. The absolute minimum value of $\chi^2$ was
equal to 232 for 104 degrees of freedom (dof), $\chi^2_{\rm red}= 2.23$.
A bin-by-bin comparison between the data and the best-fit model
is shown in Figure~\ref{north_fit}.
Parameter ranges that were obtained from the fits are presented in
Table \ref{param_ranges}. The fit was most sensitive to the pitch
angle $p$. The vertical scale of the disk MF $z_0$ and
the distance to the nearest reversal $d$ affect the fit less. There
was no significant influence of the $R_c$ parameter, so we excluded it
from further studies fixing its value at $R_c=5~$kpc. The same
concerns the parameters $z_{1(1)}^H$ and $z_{1(2)}^H$ which were fixed
at $z_{1(1)}^H=0.25$~kpc and $z_{1(2)}^H=0.4$~kpc.

We could not make a successful fit using the disk component only. The
addition of the halo component decreases $\chi^2_{\rm min}$ by a
factor of 1.5. Due to a weaker effect of the halo MF, its parameters
are estimated with larger uncertainties than the disk ones. In
addition, there is a degeneracy between the height of the halo
$z_0^H$, the amplitude of the halo field and, to some extent, the
radial size $R_0^H$. Even unrealistically large values of the halo
field of order tens of $\mu$G could give an acceptable fit if the halo
is placed at a large distance from the galactic plane where the
concentration of the electrons greatly decreases
(cf. Figure \ref{degeneracy}).  There is also a degeneracy between the
vertical scales of the GMF model (both halo and disk) and the vertical
scale of the electron density (recall that the latter was held fixed
during the fits).

The strength of the disk field corresponding to the allowed set of
models resides in the $1.8--2.8\,\mu$G range, largely concentrating
around $2.0~\mu$G. The amplitude of the allowed halo field is in the
range $2--12~\mu$G. Note that the halo field strength is constrained
by the synchrotron radio-emission from the population of Galactic
relativistic electrons \citep{Haslam1982}. These constraints favor
halo fields with the magnitude in the range $2--5~\mu$G.

Ring models of the disk field performed rather poorly in the Northern
hemisphere no matter what was the halo field, giving the best value of
$\chi^2_{\rm min}= 288$ for 101 dof, which corresponds to
$\chi_{\rm red}^2 = 2.85$ as compared to $\chi_{\rm red}^2= 2.23$ for
the best-fit model with the spiral disk. Thus, these models are
disfavored by the NVSS data, confirming the earlier finding of
Simard-Normandin \& Kronberg(1980).

Finally, our attempts to improve the quality of the fit by introducing
a non-zero halo pitch angle were unsuccessful: there was no
significant change of $\chi^2$ when non-zero pitch angle of the halo
field was varied within the range $[-10^\circ,10^\circ]$. We also could
not improve the fit by choosing the alternative model of the halo
described by Equations~(\ref{halo2}) and  (\ref{halo3}).

\begin{figure}
\begin{center}
\includegraphics[width=8cm]{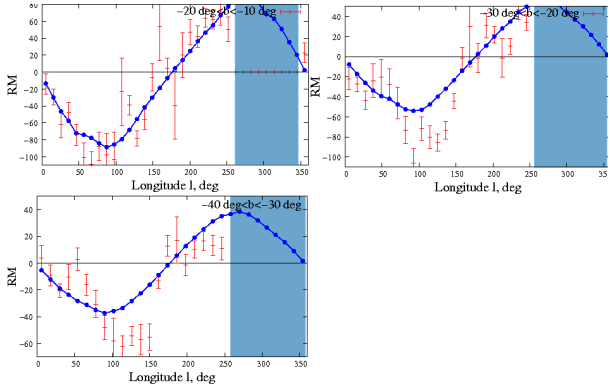}
\end{center}
\caption{Best-fit model (blue solid line) vs. NVSS data (red points)
in the Southern hemisphere. Bins not included in the fit are shaded
with light blue.}
\label{south_fit}
\end{figure}

\textit{South}. The data from the Southern hemisphere are more sparse
because of the blind spot, and show more scatter, so the fits are
generally worse. The best-fit ASS-type model gives $\chi^2 = 243$
  for $65$ dof, which corresponds to $\chi^2_{\rm red} =
  3.74$. The results for the best-fit BSS model are $\chi^2 = 235$
  for $65$ dof ($\chi^2_{\rm red} = 3.62$).
The bin-by-bin comparison between the data and the
best-fit model in the Southern hemisphere is shown in
Figure~\ref{south_fit}.

\begin{figure}
\begin{center}
\includegraphics[width=8cm]{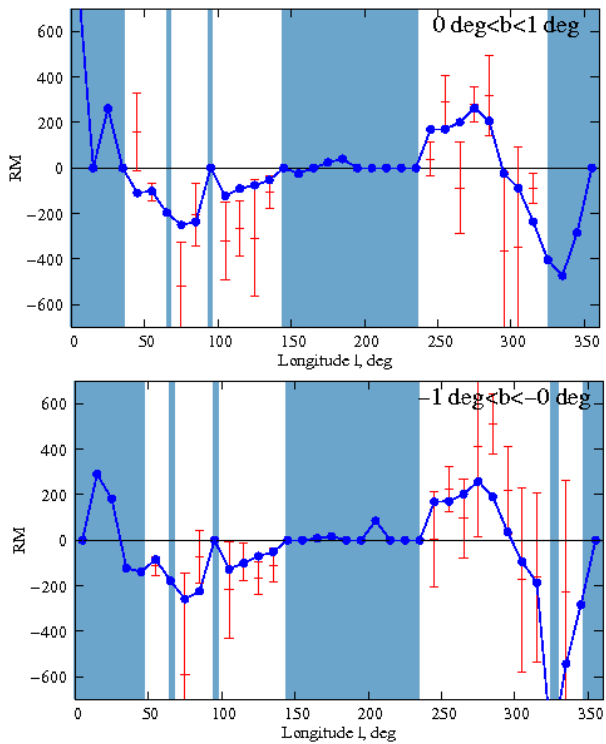}
\end{center}
\caption{Comparison between the best-fit model and the data from
  KNM11 in the strip $-1^{\circ}<b<1^{\circ}$.  Areas
  not included into the fit are shaded with light-blue.}
\label{plane_fit}
\end{figure}

{\it Disk}. Finally, we fitted the RMs from KNM11 at very low
latitudes $|b|<1^\circ$. The best-fit ASS-type model gives $\chi^2 =
30$ for $29$ dof ($\chi^2_{\rm red} = 1.03$), while the best-fit
BSS model gives $\chi^2 = 37$ for $29$ dof  ($\chi^2_{\rm red} =
1.28$).  The comparison between these data and the best-fit model is
shown in Figure~\ref{plane_fit}.
The results are fully compatible with the other fits. They also
agree with the results of \citet{Kronberg2011}. Unfortunately,
there is not enough information about the region around the
direction to the Galactic center where the discrepancy between
predictions of different models is the largest. This reduces the
discriminative power of the fit. Nevertheless, the fit was very
sensitive to the field strength in the disk, so its results impose
most stringent constraints on this parameter. Also, it was shown
by \cite{Kronberg2011} that the KNM11 data favor the pitch angle
near the disk of $-5.^\circ5$. Taking this constraint into account
allows us to exclude positive pitch angles, which is not possible
from the fits to the NVSS data alone.

In both north and south fits of the NVSS data, the reduced
$\chi^2$ values significantly exceed unity (especially in the
Southern hemisphere). This could result either from our
underestimation of errors, or from features in the data that
cannot be accounted for by our models. Visual examination of the
fits (Figures~\ref{north_fit} and \ref{south_fit}) shows that there
exists a systematic discrepancy between the data and the best-fit
models in the region $l\approx 130^{\circ}-- 220^{\circ}$ which
corresponds to  directions away from the Galactic center. The
discrepancy is present in both Northern and Southern hemispheres.

In fact, the observed variation of RM in the outer Galaxy is too
strong to be accounted for in {\em any} model that has a thick
disk, or halo with a nearly tangential field in the outer Galaxy
(i.e., small pitch angle) if the field magnitude decays away from
the Galactic center. So, either these models are completely wrong
in this region, or the large RMs are caused by some
nearby anomaly in the MF or the electron density,
which is sufficiently close to subtend a large angle of about
$\sim 90^\circ$. The latter option is supported by the fact that
the variation of RM which is responsible for this feature does not
visibly decrease away from the Galactic plane, as it would if it
were due to an additional galactic arm or another global (and more
distant) feature. Interestingly, the anomaly in the nearby region
(the so-called region A) was pointed out previously
\citep{Simard-Normandin1980}.

For an additional check of this assumption we made a separate fit of
RMs of the sources located in the direction of the inner Galaxy,
i.e., at $-90^{\circ}<l<90^{\circ}$. Excluding this apparent anomaly
should improve the fit and show the extent to which the best-fit
parameters were affected by its presence. The results of this
additional test are as follows.

\textit{North-inner}. The best fit gives $\chi^2_{\rm min}=43$ for
46 dof, considerably better than for the full sky. The allowed
values of parameters obtained from the restricted fit largely
coincide with those obtained from the full fit. This indicates
that the full fit is not dominated by the anomalous region.

\textit{South-inner}.
$\chi^2_{\rm min}=35$ for 22
dof. The quality of the fit has considerably increased as compared
to the fit in the full range of longitudes, but is still worse than in
the Northern hemisphere.

All our attempts to fit the outer Galaxy separately
($90^{\circ}<l<270^{\circ}$), even those including an extra disk
component, were unsuccessful in the sense that we could not get
$\chi^2_{\rm red}$ smaller than five. This confirms our conclusion that
it is not possible to fit the anomalous region with the GMF models of
the type considered here.

The results of individual fits are presented in
Table~\ref{param_ranges} in the form of acceptable ranges of model
parameters. It should be stressed that there are dependencies and
degeneracies among them; Table~\ref{param_ranges} only shows the
maximum variation of each parameter within the whole acceptable
set of models. In other words, not every combination of the
acceptable parameters gives an acceptable model.  Parameters of
the models which fit the data best and are in agreement with all
fits are summarized in Tables \ref{disktable}, and \ref{halotable}.

\begin{deluxetable}{ccccc}
\centering
\tabletypesize{\footnotesize}
\tablewidth{0pt} \tablecolumns{8}

\tablecaption{ \label{disktable}}
\tablehead{\colhead{}&\colhead{$p$} & \colhead{$z_0$} & \colhead{$d$}& \colhead{$B_0^D$}\\
\colhead{}&\colhead{deg} & \colhead{kpc} & \colhead{kpc}& \colhead{$\mu$G}\\
}
\startdata
ASS &
$[-5, -4]\cup [3,7]$
&$[0.7,1.2]$ &$[-0.7, -0.4]$& $[1.8,2.2]$\\
~&~&~&~&~\\
BSS &  $[-7,-3]\cup[5,8]$
&$[0.8,1.2]$&$[-0.9, -0.5]$& $[1.8,2.4]$\\
\enddata
 \tablecomments{Ranges of parameters of the disk field corresponding to
    $\sim 1\sigma$ deviation from the best fit to the NVSS data and
    compatible with the KNM11 data in the Galactic disk. Note that
    positive ranges of the pitch angle, although compatible with our
    fits, are strongly disfavored by the ``shift-and-reflection''
    arguments of KNM11. }
\end{deluxetable}

\begin{deluxetable}{cccc}
\centering
\tabletypesize{\footnotesize}
\tablewidth{0pt} \tablecolumns{8}

\tablecaption{ \label{halotable}}
\tablehead{\colhead{} & \colhead{$z_0^H$} & \colhead{$R_0^H$}& \colhead{$B_0^H$}\\
\colhead{} & \colhead{kpc} & \colhead{kpc}& \colhead{$\mu$G}\\
}
\startdata
&&North&\\
vs. ASS &[1.2, 2.4] & [6.0, 15]& [4.0, 12]\\
vs. BSS &[1.0, 2.0] & [8.0, 15] & [3.0, 12]\\
&&South&\\
vs. ASS &[1.0, 2.5]&[3.5, 15]& [1.5, 6.0]\\
vs. BSS &[1.0, 2.2]&[4.0, 15]&[2.5, 5.0]\\
\enddata
 \tablecomments{The same as Table \ref{disktable}, but for the halo
    parameters. }
\end{deluxetable}

Finally, let us comment briefly on the other tests that we have
made. First, as has been already mentioned in
Section~\ref{sec:electron-density}, we have run preliminary
simulations with the original, lower  vertical scale of electron
density. For the realistic strength of the disk field
$B_0>1.8\,\mu G$ and the original electron density, we found only
$\chi^2$ values larger than $390$ as compared to $232$ for the
best-fit case. If we  vary the magnitude of the disk field, the
value of $\chi^2$ becomes close to the best-fit value for field
strengths $\sim 0.7\,\mu G$, which is clearly unacceptable. In
conclusion, our simulations disfavor the original vertical scale
of the electron density.

Second, we have tested different combinations of symmetries (with
respect to the Galactic plane) of the halo and disk fields in
order to check whether the NVSS data alone favor the symmetric
disk and antisymmetric halo as we have chosen in our model. With
our fitting procedure which treats the Northern and Southern
hemispheres independently, no additional simulations are required
for this test. For instance, for an anti-symmetric disk field we
simply have to require that the coefficients with which the disk
contribution enters the Northern and Southern hemispheres are
opposite in sign. We could not find a satisfactory fit neither
with ASS nor with BSS anti-symmetric disk field, obtaining the
values of $\chi^2$ more than four times larger than the best-fit
value. In the case of the symmetric disk and symmetric halo, the
best fit gives $\chi^2$ 2.5 times larger than the best-fit model.
Thus, the combination of the symmetric (with respect to $b$) disk
and antisymmetric  (with respect to $b$) halo is favored by the
NVSS data.

\section{Conclusions}
\label{sec:conclusions}

The latest Faraday rotation data show an unmistakable coherent
component of the RMs over the sky. The RMs are correlated in sign and value over the largest angular
scales. Moreover, the distribution of RMs is correlated with the
Galactic plane and the direction to the Galactic center. This
strongly indicates the Galactic origin of this coherence and
provides firm evidence for the existence of a regular GMF.

Apart from the Galactic plane region $|b|\lesssim 10^\circ$, the
global pattern of the RMs has different symmetry
with respect to the Galactic plane in the inner and outer Galaxy:
symmetric outside and anti-symmetric inside (see Figure~
\ref{fig:bin_nvss}). This change of the symmetry cannot be
achieved with a single (symmetric or antisymmetric) component of
the GMF. At least two components---disk and halo---are required (Figure~\ref{sketch_model}). Since the study of the
MF in the disk shows that it does not change sign
across the Galactic plane itself, the combination of the symmetric
disk field and the anti-symmetric halo field is favored. We have
demonstrated that such a combination, with properly chosen
amplitudes of components, reproduces the observed pattern of RMs
very well.

Several further conclusions can be made.
\begin{itemize}
\item[1.] The cross check between the NVSS and the smaller KNM11 set
  shows that the quality of the NVSS is sufficient for our purposes
  all over the sky except, in the $|b|<10^{\circ}$ band around the Galactic
  plane.
\item[2.] There is a feature  in the observed RMs near the direction to the Galactic anti-center that cannot be
  accounted for by models of the type considered in this paper (nearly
  tangential field decaying away from the Galactic center). This
  feature may be due to some unspecified local structure.
\item[3.] A ring model of the GMF is disfavored as compared to the
  model with logarithmic spiral arms. In the former case the best
  value of $\chi^2_{\rm red}$ is a factor of $\geq 1.3$ larger than in
  the latter.
\item[4.] Our fits favor the increased vertical scale of the electron
  density distribution as proposed by \cite{Gaensler2008}.  We were
  unable to find any fit with the original vertical scale of NE2001 model
  and strong disk field. This is in agreement with the results of
  \citet{Sun2010}.
\item[5.] Our fits favor somewhat smaller values of the pitch angle
  $p$ than have been discussed in the literature previously. These smaller
  values agree with the recent results of \cite{Kronberg2011} where
  the reference value of $p=-5.^\circ5$ was obtained by a totally
  different method.
\end{itemize}

Our fitting procedure could not discriminate between the ASS and BSS
models because a major contribution to the extragalactic source RMs
(especially at $|b| > 15^{\circ}$) comes from the near vicinity of the
solar system; in this region both models behave quite similarly. A
more detailed analysis of the Galactic plane region may be more
sensitive to the differences between the ASS and BSS models.

Two ``benchmark'' models (one with the ASS and one with the BSS
disk field) are presented in Table~\ref{tab:benchmark}. These
models fit the RM data well and have values of parameters favored
by other studies (not too small pitch angle $p$, not too strong
halo field and the magnitude in the vicinity of the Earth close to
$2~\mu$G).  They can be used, e.g., for propagation of cosmic rays
through the Galaxy. At ultra-high energies, an accurate three-dimensional
description of the Galactic magnetic field may be crucial for
identification of extragalactic cosmic ray sources.

\begin{deluxetable}{lcc}
\centering
\tabletypesize{\footnotesize}
\tablewidth{0pt} \tablecolumns{8}

\tablecaption{Benchmark model parameters. \label{tab:benchmark}}
\tablehead{\colhead{}&\colhead{ASS} & \colhead{BSS} \\}
\startdata
Disk: &&\\
$p$ & $-5^\circ$ & $-6^\circ$\\
$z_0$ & $1.0$~kpc &  $1.0$~kpc \\
$d$ & $-0.6$~kpc &  $-0.6$~kpc \\
$B_0$ & $2.0~ \mu$G &  $2.0~ \mu$G \\
$R_c$ & $5.0$  kpc &  $5.0$  kpc \\
\\
Halo (North): &&\\
$z_0^H$ & $1.3$~kpc &  $1.3$~kpc \\
$R_0^H$ & $8$~kpc & $8$~kpc \\
$B_0^H$ & $4\mu$G &  $4\mu$G \\
$z_{1(1)}^H$ & $0.25$~kpc &  $0.25$~kpc \\
$z_{1(2)}^H$ & $0.4$~kpc &  $0.4$~kpc \\
\\
Halo (South): &&\\
$z_0^H$ & $1.3$~kpc &  $1.3$~kpc \\
$R_0^H$ & $8$~kpc & $8$~kpc \\
$B_0^H$ & $2\mu$G &  $4\mu$G \\
$z_{1(1)}^H$ & $0.25$~kpc &  $0.25$~kpc \\
$z_{1(2)}^H$ & $0.4$~kpc &  $0.4$~kpc \\
\enddata

\end{deluxetable}

In this paper we used only the Faraday RMs of
extragalactic radio sources. More precise understanding of the
GMF can be achieved by taking into account of
other data, also sensitive to the MF, such as RMs of
pulsars, measurements of the galactic synchrotron radiation, etc.
This would permit better constraints on the unknown quantities
such as the three-dimensional electron density profiles and other properties of
the ISM. It will also help to resolve some degeneracies inherent
in our models.



\section*{Acknowledgments}
The authors are grateful to D.D. Sokoloff and JinLin Han for
fruitful discussions. Also, we thank an anonymous
referee for careful reading of the manuscript and valuable
remarks. This work is supported by IISN project No. 4.4509.10 and
the Natural Sciences and Engineering Research Council of Canada
(NSERC), the Australian Research Council, the U.S. Department of
Energy, and the use of NASA's Astrophysics Data System.

\appendix
\section{Error assignment. Excluded bins and bins with manually
assigned errors}
\label{ExcludedBins}

In our fitting procedure, each bin included in the analysis was
characterized by an average RM and its error $\delta$.
Here we describe the algorithm by which these errors were assigned
on a bin-by-bin basis.

First, we estimated the intrinsic dispersion of RMs of extragalactic
sources. To this end we investigated the regions on the sky where the
influence of the GMF is expected to be suppressed. Since RMs are
proportional to the MF component that is parallel to the
line of sight, these are the regions around the Galactic
poles. Considering the sources with $|b|>80^{\circ}$ we found that the
dispersion of RMs of individual sources in these regions was about
$\sigma_{\rm qso}= 15\,\text{rad\,m}^{-2}$. Comparing to
Figure~\ref{figure_nvss-vs-Kr} we see that the dominant contribution to
this number comes from the measurement errors. In the error assignment
algorithm, we adopted the value of $\sigma_{\rm qso}$ as an indicator
of whether a given bin is perturbed by effects other than the
random scatter of sources.

If a bin showed the dispersion of RMs $\sigma_i$ smaller than
$\sigma_{\rm qso}$ we considered it unperturbed and assigned its error
on a statistical basis. Since the average of a sample of $N$ elements
varies around the true average with the width $\sigma/\sqrt{N}$, where
$\sigma^2$ is the variance of the distribution, we assigned to such bins
an error of $\delta_i = \sigma_{\rm qso}/\sqrt{N_i}$, $N_i$ being the
number of sources in the bin. In practice, this situation occurs only
for high-latitude bins.

In the opposite case, i.e., when the dispersion $\sigma_i$ was larger
than $\sigma_{\rm qso}$, we considered the bin as perturbed by
coherence effects. These may be due to, e.g., random MF
with the coherence length comparable to the bin size, fluctuations of
electron density, strong variation of the regular component across the
bin, etc. In this case the statistical approach is not correct, and a
precise determination of the error is difficult. We have assumed that
the error in such bins is some fraction of the bin variance
$\sigma_i$, which we have adopted to be $1/3$, so that the error in
perturbed bins was taken $\delta_i = \sigma_i/3$. We cannot rigorously
derive this coefficient (although it can be motivated by a toy model
presented below). Since this uncertainty propagates directly into the
value of $\chi^2$, it results in some uncertainty of the absolute
normalization of the latter. Note, however, that the ratios of
$\chi^2$ are less affected by this uncertainty.

The following toy model shows how the proportionality between the
error and the dispersion in the bin may come about. Let the random part of
the MF consist of cells of a typical size $d\sim 100$~pc,
such that their contributions to RM have zero mean and the dispersion
$\sigma_0$. If the cells were statistically independent, individual
sources would have an RM dispersion given by $\sigma = \sigma_0
\sqrt{N_{\rm tot}}$, where $N_{\rm tot}$ is the total number of cells
along the line of sight (we have neglected the internal
scatter). However, for sources in one bin a number $N_{\rm coh}$ of
nearby cells contributes {\em coherently}. If $N_{\rm coh}\ll N_{\rm
  tot}$ this will not change the dispersion $\sigma$ over the bin. The
error, on the contrary, will be dominated by the coherent
contribution, $\delta = \sigma_0\sqrt{N_{\rm coh}} =
\sigma\sqrt{N_{\rm coh}/N_{\rm tot}}$. Thus, the error and the
dispersion are proportional to one another with the coefficient
$\sqrt{N_{\rm coh}/N_{\rm tot}}$. The latter coefficient is purely
geometrical. Taking $N_{\rm coh}\sim 1/\theta$, where $\theta$ is
the bin size, and $N_{\rm tot}\sim L/d$, where $L$ is the length
of the trajectory inside GMF, we find $\delta \sim \sigma/3$.
Interestingly, similar conclusions (proportionality and a similar
coefficient) are obtained by considering a perturbation caused by
a close group of a few sources having similar RMs substantially
different from the rest of the sources in the bin.

As a last step, all bins were subject to visual scrutiny, with the
result that some of the bins were completely removed from the
analysis. These are the bins showing obvious pathologies like
small number of sources or non-uniform distribution across the bin
(e.g., some bins at the boundary of the NVSS blind spot), compact
aggregates of sources with RMs drastically different from the
surrounding value, etc. Finally, some bins with such pathologies
were assigned large errors (2--3 times larger than prescribed by
the above algorithm) by hand. This was done in cases where the
corresponding bin might serve as a good discriminator between
models, even with a large error. A constant error of
$40\,\text{rad\,m}^{-2}$ was used in this case. This value was
chosen in such a way that these bins do not contribute
significantly into the value of $\chi^2$, at least near the
best-fit point. We have checked that the results of the fits do
not depend on the precise choice of these errors. The number of
such bins is about 10\%; they are easily identifiable in
Figures~\ref{figure_bands_north} and \ref{figure_bands_south}.

Here is the complete
list of excluded bins and bins with manually assigned errors.

{\it Northern hemisphere.}  An error of $40\,\text{rad\,m}^{-2}$ was
assigned to the bins in the region $30^{\circ}<l<60^{\circ},~
10^{\circ}<b<40^{\circ}$ (local structures described in
\citep{Wolleben2010}); to the bins in the area
$-115^{\circ}<l<-90^{\circ},~ 20^{\circ}<b<30^{\circ}$ (a prominent
structure with large positive RMs, possibly related with the Gum
nebula); to the bin located at $(15^{\circ},25^{\circ})$ (strong
non-uniformity: only 12 sources are located at
$10^{\circ}<b<15^{\circ}$, while 38 at $15^{\circ}<b<20^{\circ}$).
Excluded from the analysis were the  bins in the area
$-10^{\circ}<l<30^{\circ},~ 20^{\circ}<b<30^{\circ}$ (unresolved
structure in RM and a large positive spot at
$(-10^{\circ},25^{\circ})$); bins in the area
$70^{\circ}<l<90^{\circ},~ 10^{\circ}<b<20^{\circ}$ (large negative
spot centered at $(80^{\circ},15^{\circ})$; bin affected by  Gum
nebula at ($l=-100^{\circ}, b<20^{\circ}$); the bin centered at
$(-30^{\circ},15^{\circ})$ (strongly non-uniform distribution).

\noindent
\textit{Southern hemisphere.} An error of 40~rad~m$^{-2}$ was assigned
to the bin $(105^{\circ},-15^{\circ})$
 (local structure resulting in positive excess ); to the bin
$(165^{\circ},~-15^{\circ})$ (positive
spot); to the bin centered at (180$^{\circ}$, --15$^{\circ}$) (negative
spot near the anti-center direction); and to the bin
$(170^{\circ},-25^{\circ})$ (large positive
spot).

\newpage

\section{Supplementary Tables and Pictures}
\label{Tables}
\begin{deluxetable}{cccccccccc}
\centering
\tabletypesize{\footnotesize}
\tablewidth{0pt} \tablecolumns{8}

\tablecaption{Ranges of disk and halo parameters of the acceptable
  models \label{param_ranges}}
\tablehead{\colhead{}&\colhead{$\chi^2$} & \colhead{$\chi^2_{\rm red}$} &\colhead {$p$}&
\colhead{$z_0$} &
\colhead{$d$} & \colhead{$\alpha$} & \colhead{$z_0^H$} & \colhead{$R_0^H$} & \colhead{$\beta$} \\
\colhead{}  &\colhead{}  & \colhead{} & \colhead{deg}  & \colhead{kpc} & \colhead{(kpc)}  & \colhead{} & \colhead{kpc}&\colhead{kpc}&\colhead{}\\
}
\startdata
\\
&  & &  && North &  & & &\\
\\
   A& 240 &2.30 &[-10, -3] &[0.5, 1.3] & [-1.1, -0.3]& [0.9, 2.0] &
   [1.0, 1.4] & [4, 15] & [0.4, 0.7] \\
   B& 232&2.23& [-6, -10] &[0.5, 1.3] & [-1.1, -0.5] & [0.9, 1.5] &
   [1.0, 1.4] & [6, 15] & [0.3, 1.0] \\
\\
\tableline
\\
 &  & & & &South  & &  & &\\
\\

   A&243 &3.74 & [-10, 10] & [0.5, 1.5] & [-1.1, 1.1] & [0.9, 2.0] &
   [1.0, 3.5] & [3.5, 15] & [0.2, 2.0] \\
  B& 235&3.62& [-10, 10] & [0.5, 1.1] & [-1.0, 1.0] & [0.2, 2.0] &
  [1.0, 3.5] & [3.5, 15] & [0.2, 2.0] \\
\\
\tableline
\\
 &  & & &  &Plane  & & &  &\\
\\

  A&30 &1.03 &[-7, 10] & [0.5, 1.4] & [-1.4, 1.4] & [0.9, 1.1]& & &\\
   B& 37&1.27&
[-8, -3]$\cup$[5, 8] &[0.5, 1.5] & [-0.8, 1.2] & [1.0, 1.2] & & &\\
\\
\tableline
\\
 &  & &  &  &North-inner  &  & & &\\
\\
  A&43 &0.93 & [-5, 6] & [0.5, 1.2] & [-1.0, 0.9] & [0.9, 3.0] & [1.2,
    2.4] & [6.0, 8.0]& [0.4, 1.2] \\
   B& 45&0.98&[-7, 9] & [0.5, 1.4] & [-0.6, 1.1] & [0.9, 1.1] & [1.0,
     2.0] & [8.0, 15] & [0.4--1.2] \\
\\
\tableline
\\
 &  & & & &South-inner &  & & & \\
\\
A&40 &1.81 &[-7, 11] & [0.5, 1.3] & [-0.2, -0.9] & [0.9, 3.5]
& [1.0, 2.5] & [8.5, 15]& [0.1, 0.6] \\
   B& 35&1.59&[-5, 9] &[0.5, 1.2] & [-0.3, -0.8]& [0.9, 1.2]&[1.2,
     2.2] & [3.5, 5.5] & [0.2, 0.5] \\
\enddata

\tablecomments{A and B denote ASS and BSS models, correspondingly. The
parameters $\alpha$ and $\beta$ determine the strengths of the disk
and halo components as explained in
Sect.~\ref{sec:fitting-procedure}. }
\end{deluxetable}

\begin{figure}
\begin{center}
\includegraphics[width=8cm]{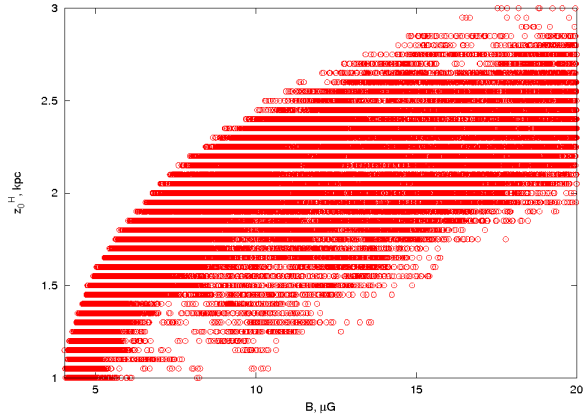}
\end{center}
\caption{Height of the halo $z_{\rm{0}}^H$ vs. its strength
  $B_{\rm{0}}^H$. Each dot represents an acceptable model. The
  degeneracy could be easily seen: a very strong halo field gives an
  acceptable fit provided the vertical scale  $z_{\rm{0}}^H$ is high.
 \label{degeneracy}
}
\end{figure}

\begin{figure}
\begin{center}
\includegraphics[width=10cm]{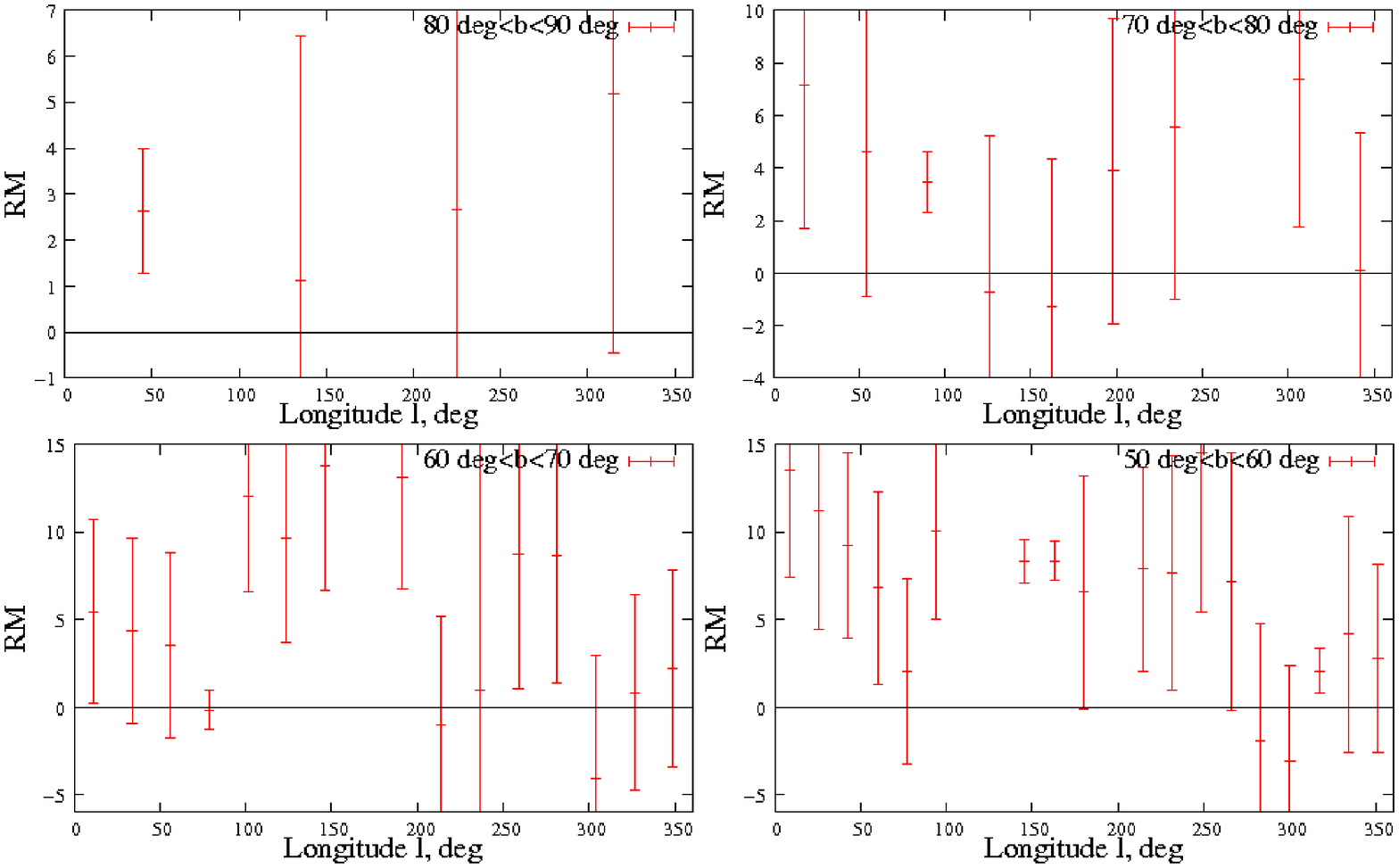}
\end{center}
\caption{NVSS data for $b>50^{\circ}$. The bias towards positive values of
  RMs, perhaps due to the small vertical component $B_z$, could be
  seen. } \label{northern_cap}
\end{figure}

\begin{figure}
\begin{center}
\includegraphics[width=10cm]{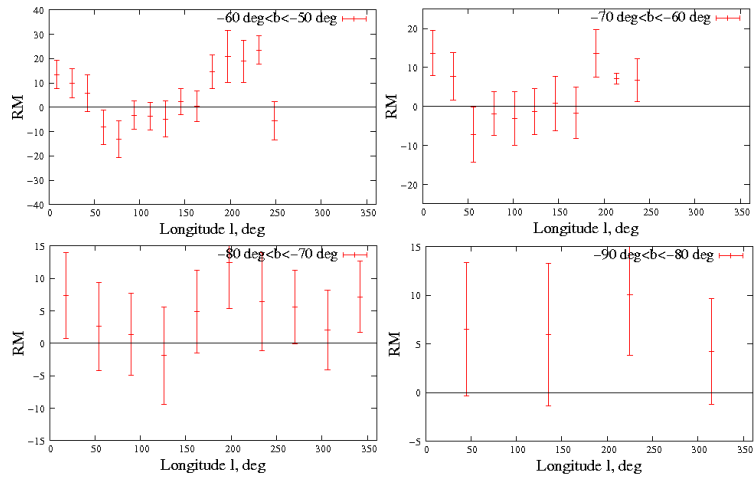}
\end{center}
\caption{NVSS data for $b<-50^{\circ}$. Bias towards positive values
  of RMs could be seen here as well. }
\label{southern_cap}
\end{figure}

\newpage



\end{document}